\documentclass[pra,twocolumn,aps, nofootinbib]{revtex4-1}

\usepackage{graphicx}
\usepackage[colorlinks=true,linkcolor=blue,citecolor=red,urlcolor=blue]{hyperref}
\usepackage{amsmath, amssymb}
\usepackage{bm}
\usepackage{comment}
\usepackage{color}
\usepackage{soul,xcolor}
\usepackage{url}

\newcommand{\be}{\begin{equation}}
\newcommand{\ee}{\end{equation}}

\begin{document}
\title{Polarisation and Transparency of Relativistically Rotating Two-Level Atoms} 
\author{Calum Maitland$^{1, 2*}$, Matteo Clerici$^{3}$, Fabio Biancalana$^{1}$}
\email{cm350@hw.ac.uk, f.biancalana@hw.ac.uk}
\affiliation{$^1$Institute of Photonics and Quantum Sciences, Heriot-Watt University, Edinburgh EH14 4AS, UK\\
$^2$School of Physics and Astronomy, University of Glasgow, Glasgow G12 8QQ, United Kingdom\\
$^3$School of Electronic and Nanoscale Engineering, University of Glasgow, Glasgow G12 8LT, UK}

\begin{abstract}
Electromagnetism and light-matter interaction in rotating systems is a rich area of ongoing research. We study the interaction of light with a gas of non-interacting two-level atoms confined to a rotating disk. We numerically solve the optical Bloch equations to investigate the how relativistic rotation affects the atoms' polarisation and inversion. The results are used to predict the steady-state stimulated emission seen by an observer at rest with the optical source in the laboratory frame. Competing physical effects due to time dilation and motion-induced detuning strongly modify solutions to the Bloch equations when the gas's velocity becomes relativistic. We account for the non-inertial motion by including acceleration-dependent excitation and emission rates, arising from a generalised Unruh effect. The effective thermal vacuum resulting from large accelerations de-polarises the gas while driving it towards population inversion, negating coherent driving due to the external light source. The results illustrate the intuitive, special-relativistic approach of assigning instantaneously comoving frames to understand non-inertial motion's influence when only local fields are physically significant.  
\end{abstract}

\maketitle
\section{Introduction}
The physical theory of relativistic rotation has been in development for over a century since Einstein's conception of special relativity, and was a source of inspiration for his general theory. The fact that publications on this subject continue to appear in the literature up to the present  \cite{Nikolic2000, Rizzi2004, Kassner2012} shows it is a complex and active area of research. Even the simple idea of a uniformly rotating frame runs into conceptual difficulties as there is a radius beyond which comoving objects exceed the speed of light, the so-called light cylinder  \cite{Eksi2005}. Several authors constructed ``relativistic rotational transforms'' in an attempt to circumvent this problem  \cite{Franklin1922, Takeno1952, Post1967, Strauss1974, NouriZonoz2014}.

The observation that light can be affected by the motion of media it travels through predates Maxwell's equations. Fizeau demonstrated the Fresnel drag of light by moving water in 1851, showing by interference that light's velocity increases/decreases when propagating with/against the fluid flow  \cite{Anderson1994}. After the advent of relativity and Maxwell's theory of electromagnetism, Minkowski developed a general framework for studying light in moving media by Lorentz transforming Maxwell's equations to the comoving frame of the material  \cite{bolotovskii}. This approach leads to Minkowski's constitutive equations, which relate the material excitation fields $D$ and $H$ to the electric $E$ magnetic $B$ fields in a medium moving at fixed velocity relative to the observer  \cite{Minkowski1908, Ivezic2012}. Complications arise in anisotropic  \cite{Tai1964} and inhomogeneous  \cite{Tanaka1972} media. More recently, Leonhardt and Piwnicki  \cite{piwnicki} have examined the problem for slowly moving media from a geometrical optics perspective, expanding on Gordon's original idea  \cite{Gordon1923} that moving dielectrics may act as effective gravitational fields for light. 

A key result of special relativity is that Maxwell's equations are Lorentz invariant, meaning their mathematical structure does not change when transforming between different inertial frames (though the fields themselves may). However, a uniformly rotating frame is non-inertial as comoving observers experience a centripetal acceleration. As Lorentz transforms are only well-defined between inertial frames, it is not immediately obvious how to relate spacetime coordinates in the comoving frame to the laboratory frame (which sees the material rotating). Ehrenfest's paradox  \cite{Ehrenfest1909} exemplifies this problem for a disc with radius $R$ when measured at rest, rotating with angular velocity $\Omega$ relative to a stationary observer. The observer sees the lengths of bodies moving towards them shortened by a factor $\gamma = 1/\sqrt{1 - {\left(\Omega R / c \right)}^2}$, where $c$ is the speed of light, due to length contraction. They would conclude that the circumference of the disc rotating underneath them is $2 \pi R' = 2 \pi R/\gamma$. This implies a disc of radius $R'$, however the radial axis between the disc rim and the origin is perpendicular to the rotation and so should be $R$ in both the comoving and observer frames. The resolution of this contradiction comes from the impossibility of synchronising clocks around a rotating ring  \cite{Gron1975}  \cite{Gron2007}. 

For the most part, experimental work in this area appears to have been limited to studies in simple dielectric materials, typically at slow medium velocities.  The Wilson-Wilson experiment  \cite{Wilson1913} (repeated and verified by Hertzberg \textit{et.al.}  \cite{Hertzberg2001}) showed that rotating a dielectric cylindrical shell, in the presence of a static magnetic field oriented along its axis, induced a electric field between its inner and outer surfaces. The strength of the electric field agreed with the result predicted Einstein and Laub  \cite{Einstein1908} using Minkowskii's constitutive relations, as applied to the cylinder's rotational motion rather than inertial translation as they were originally derived for. As noted in ref.  \cite{Canovan2010}, applying theory valid for special relativistic inertial motion to non-inertial rotation was considered na\"{i}ve and controversial. A series of publications attempted to find a more satisfactory explanation  \cite{Pellegrini1995, Burrows1997, Weber1997, Ridgely1998, McDonald2008}. The most recent contribution on this problem by Canovan and Tucker  \cite{Canovan2010} derived a general result for the electric and magnetic fields in the rotating cylinder using differential forms, which in the limit of a non-relativistic rotational velocity recovers the solution found by the experiments  \cite{Wilson1913, Hertzberg2001}. This suggests that when tranforming electromagnetic fields to reference frames of non-inertial observers, the method of Lorentz transforming the fields from the inertial frame to an another inertial frame, instantaneously comoving with the non-inertial observer, is at least a reasonable approximation. However, Mashhoon's articles \cite{Mashhoon2008, Mashhoon2011, Mashhoon2012} demonstrate that measurements by non-inertial observers are necessarily nonlocal and have to account for the field's history by integrating over the observer's past wordline. The form of the kernel for this integral has yet to be determined \cite{Mashhoon2012b}, but the nonlocal contribution should vanish in the limits of either weak accelerations or vanishing wavelength (the eikonal/ray optics limit) \cite{Mashhoon2012}. We posit that either when these limits hold or the physics under consideration is essentially local and approaching a steady state, the affects of uniform, relativistic non-inertial motion can be described sufficiently by both using instantaneously comoving frames (as suggested by Einstein and other authors \cite{Desloge1987}) and accounting for the accelertion-modified vacuum \cite{Audretsch1995}. 

In this work, we will focus on the optical Bloch equations for a gas of non-interacting two-level atoms confined to a disc, rotating at a constant angular frequency $\Omega$ relative to the source of a coherent, monochromatic light field propagating at normal incidence through the disc. To our knowledge such a system has not been studied before, yet the Bloch equations give a straightforward and quantitatively accurate model of light interacting with simple atomic matter. As such they are fairly simple to study in a corotating frame. Since the exact treatment for the nonlocal aspects of non-inertial relativity has yet to be determined, we neglect it here under the assumption that it contributes a small, transient correction that has little effect on the system's long-term dynamics.  We first solve the full Bloch equations numerically in a simplified rectilinear problem where the relativistic physics are completely unambiguous, \textit{i.e.} the atomic comoving frames are inertial and connected to the laboratory frame by Lorentz transforms. We find reasonable agreement between this result and the analytical steady-state solution to the Bloch equations in the rotating wave approximation. We then show how the solution of this problem is analogous to the solution of the Bloch equations for the rotating gas, under the assumption that Lorentz transforms to frames instantaneously comoving with the atoms are sufficient to describe this non-inertial problem.  For small, rapidly rotating discs which experience greater acceleration, the solution is modified compared to the inertial case as the atoms feel a finite vacuum temperature. Hence we extrapolate the potential stimulated emission intensity seen from a laboratory observer's perspective based on steady-state solutions for the rotating disc's polarisation under different pumping regimes.\\

\section{Linear Flow Problem}

As a preliminary step we will demonstrate the essential physics in a Cartesian analogue for the rotating disc, a sheet of gas flowing along a channel in the $x$ direction with a linear velocity gradient $\Omega$ in the transverse $y$ direction; $\mathbf{v}(y) = \Omega y \hat{\mathbf{x}}$. Light propagates orthogonally to this channel in the $z$ direction. We choose the flow velocity gradient $\Omega$ and the width of the gas channel such that the atoms' speed varies from zero at the inner edge $y=0$ to $99\%$ of the speed of light at the outer edge  $y=y_{max}$, \textit{i.e.} $\Omega y_{max} = 0.99 c$. We take $y_{max} = 1$ in some arbitrary length scale. In this case an inertial frame comoving with the gas can be identified for each position $y$, which can related to the laboratory frame using standard Lorentz transformations:
\begin{equation} \label{simpleLorentz}
\begin{split}
t' &= \gamma(y) \left( t - v(y) x / c^2 \right)  \\
x' &= \gamma(y) \left( x - v(y) t \right)\\ 
y' &= y
\end{split}
\end{equation} 
where here $\gamma(y) = 1/\sqrt{1 - {\left(\Omega y / c \right)}^2}$. We assume the atoms have a simple electric dipole moment $\mu$ and no magnetic moment in their rest frame (a good approximation for noble gas atoms with all electronic shells full), do not interact with each other or the channel boundaries and that their configuration does not allow for collective optical excitations. Both the flow and the light are homogeneous in the horizontal coordinate $x$, considered infinite in extent. The electric and magnetic field vectors from the continuous-wave laboratory frame source are $\mathbf{E} = E_0 \mathbf{\hat{y}}$, $\mathbf{B} = (E_0 /c) \mathbf{\hat{x}}$ given $E_0 = A \exp{\left(-i \omega_E t \right)}$ with $A$, $\omega_E$ constant. Standard Lorentz transformations then give the fields seen in the atoms' comoving frame:  
\begin{equation}
\begin{split}
\mathbf{E'}_\perp &= \gamma \left(\mathbf{E}_\perp + \mathbf{v} \times \mathbf{B} \right)\\
\mathbf{B'}_\perp &= \gamma \left(\mathbf{B}_\perp - \frac{1}{c^2}\mathbf{v} \times \mathbf{E} \right)\\
\mathbf{E'}_\parallel &= \mathbf{E}_\parallel \\
\mathbf{B'}_\parallel &= \mathbf{B}_\parallel \\
\end{split}
\label{eq:SRfields}
\end{equation} 
where the subscripts $\perp$ and $\parallel$ indicate vector components perpendicular and parallel to the gas velocity respectively; this simplifies to 
\begin{equation} \label{simpleFieldLorentz}
\begin{split}
\mathbf{E'} &= \gamma(y) E_0 \mathbf{\hat{y}} = \gamma(y)\mathbf{E} \\
\mathbf{B'} &= \frac{E_0}{c} \left(\mathbf{\hat{x}} - \gamma(y) \frac{\Omega y}{c} \mathbf{\hat{z}} \right) = \mathbf{B} - \gamma(y) \frac{\Omega y}{c^2} E_0 \mathbf{\hat{z}}.  \\ 
\end{split}
\end{equation} 
This specifies the electromagnetic field in the frame corotating with the atoms; all that remains is to specify how the atoms interact with this field. In the comoving frame, the atoms are at rest  and their interaction with light only depends on their electric dipole moment, so the magnetic field $\mathbf{B'}$ is irrelevant. The electric field orientation is the same in both frames, however its strength is boosted by $\gamma(y)$ in the comoving frame. Further the optical frequency $\omega_E' = \omega_E / \gamma(y)$ seen by the atoms will vary with $y$ due to the (transverse) relativistic Doppler shift. This results in a detuning across the gas channel's transverse direction $y$, which may be comparable to the transition frequency of the two-level system. Hence the rotating wave approximation is not necessarily valid and more general Bloch equations are required. The two level system is described by the state $| \psi(t') \rangle = c_g(t') |g \rangle + c_e(t') |e \rangle$ given that the ground state is $|g \rangle$ and the excited state is $|e \rangle$ and $c_g, c_e$ are the complex amplitudes of each in the atomic state at time $t'$. This can be expressed in terms of a density matrix 
%% note = {URL: http://community.dur.ac.uk/thomas.billam/PreviousNotes_MPAJones.pdf, accessed 13/07/18},
%note = {URL: https://www.physik.hu-berlin.de/de/nano/lehre/copy_of_quantenoptik09/Chapter7, accessed 13/07/18},
\begin{equation} \label{eq: densitymatrix}
\hat{\rho}  = \begin{pmatrix}
\rho_{gg}  &  \rho_{ge} \\
\rho_{eg}  &  \rho_{ee} \\
\end{pmatrix} = 
\begin{pmatrix}
c_g {c_g}^*  &  c_g {c_e}^*  \\
c_e {c_g}^*  &  c_e {c_e}^* \\
\end{pmatrix}
\end{equation}
If spontaneous emission is accounted for by an excited state inverse lifetime $\Gamma$ and the Hamiltonian for the two-level atom interacting with the light field is
\begin{equation} \label{eq: 2levelH}
\hat{H}(t') = \frac{1}{2}
\begin{pmatrix}
0  &  \langle g | \mathbf{\mu}\cdot \mathbf{E'(t')} | e \rangle \\
\langle e | \mathbf{\mu}\cdot \mathbf{E'(t')} | g \rangle  &  2 \hbar \omega_0 \\
\end{pmatrix}
\end{equation}
the evolution of the density matrix is given by Liouville's equation \cite{DurhamNotes}
\begin{equation} \label{eq: Liouville}
\frac{d}{dt'} \hat{\rho}  = \frac{i}{\hbar} [ \hat{\rho} , \hat{H} ] - 
\begin{pmatrix}
-\Gamma \rho_{ee}  &  \frac{\Gamma}{2}\rho_{ge} \\
\frac{\Gamma}{2} \rho_{eg}  &  \Gamma \rho_{ee} \\
\end{pmatrix} \\
\end{equation}
Introduce the Bloch vector, defined by \cite{HumboltNotes}
\begin{equation} \label{eq: BlochVec}
\mathbf{b} = \begin{pmatrix}
b_1 \\
b_2 \\
b_3 \\
\end{pmatrix} = 
\begin{pmatrix}
2  Re( \rho_{eg}) \\
2 Im( \rho_{eg} ) \\
\rho_{ee} - \rho_{gg}
\end{pmatrix}
\end{equation}
This is related to the inversion (population difference between excited and ground atomic states) 
\begin{equation} \label{eq: W}
W = b_3
\end{equation}
and dimensionless atomic polarisation (coherence) 
\begin{equation} \label{eq: P}
P  = \left({b_1} + i{b_2} \right)/2. 
\end{equation}
Defining the comoving Rabi frequency by  
\begin{equation} \label{eq: Rabi}
\begin{split}
\Omega_R'(y)  &= \frac{1}{\hbar} \langle g | \mathbf{\mu}\cdot |\mathbf{E'}| | e \rangle \\
&=  \langle g | \mathbf{\mu}\cdot \mathbf{\hat{y}} | e \rangle \frac{A \gamma(y)}{\hbar} = \gamma(y) \Omega_R
\end{split}
\end{equation}
($y$-dependent due to the Lorentz boost of the electric field), the optical Bloch equations for these variables are derived from eq. \ref{eq: Liouville}, making use of the constraints $\rho_{ee} + \rho_{gg} = 1$ and $\rho_{eg} = {\rho_{ge}}^*$
\begin{equation} \label{eq: generalBloch}
\begin{split}
\dot{b_1} = &- \omega_0 b_2 + \Omega_R'(y) \sin{(\omega_E'(y) t')}b_3  - \frac{\Gamma}{2} b_1\\
\dot{b_2} = &\omega_0 b_1 - \Omega_R'(y) \cos{(\omega_E'(y) t')} b_3  - \frac{\Gamma}{2} b_2\\
\dot{b_3} = &-\Omega_R'(y) \sin{(\omega_E'(y) t')}b_1 + \Omega_R'(y) \cos{(\omega_E'(y) t')} b_2\\
 &-\Gamma \left( b_3 + 1 \right)
\end{split}
\end{equation} 
\begin{figure}[h] 
\centering
\includegraphics[width=0.9\linewidth]{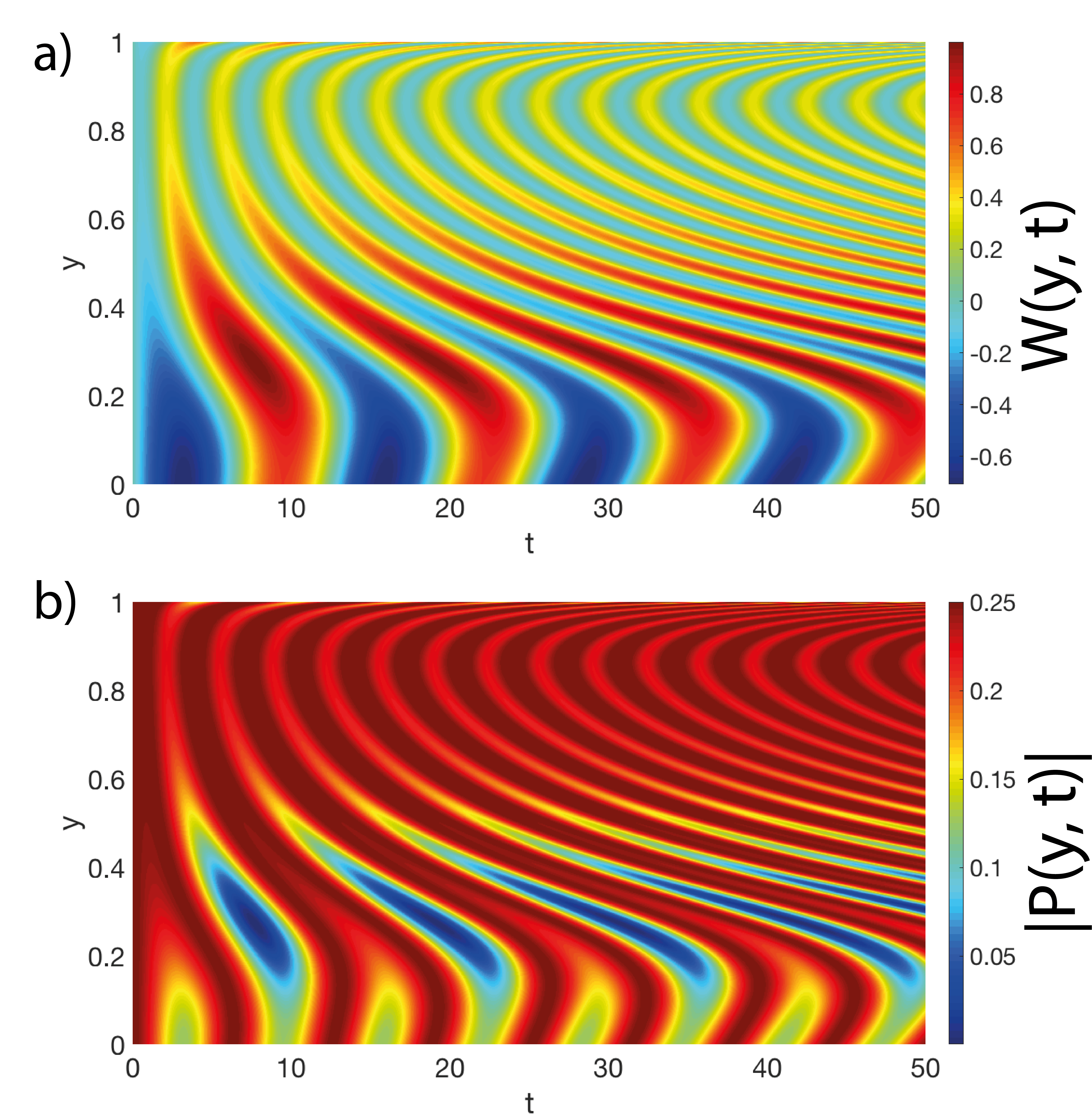}
\caption{\small{Inversion $W$ \textbf{a)} and polarisation magnitude $|P|^2$ \textbf{b)} as defined by equations \ref{eq: W} and \ref{eq: P} respectively, induced by a coherent, homogeneous source in a line of atoms flowing relative to both the source and observer with velocity profile $\mathbf{v}(y) = \Omega y \hat{\mathbf{x}}$. One time unit corresponds to one sixth of the stationary (laboratory frame) Rabi oscillation period $\tau_R \equiv 2 \pi /\Omega_R$. Other frequency parameters are chosen such that the interaction is resonant at $y=0$ and in a weak driving regime: $\omega_0 = \omega_E = 10\Omega_R$.}}  
\label{fig: cartflow}
\end{figure}
\begin{figure}[h] 
\centering
\includegraphics[width=0.8\linewidth]{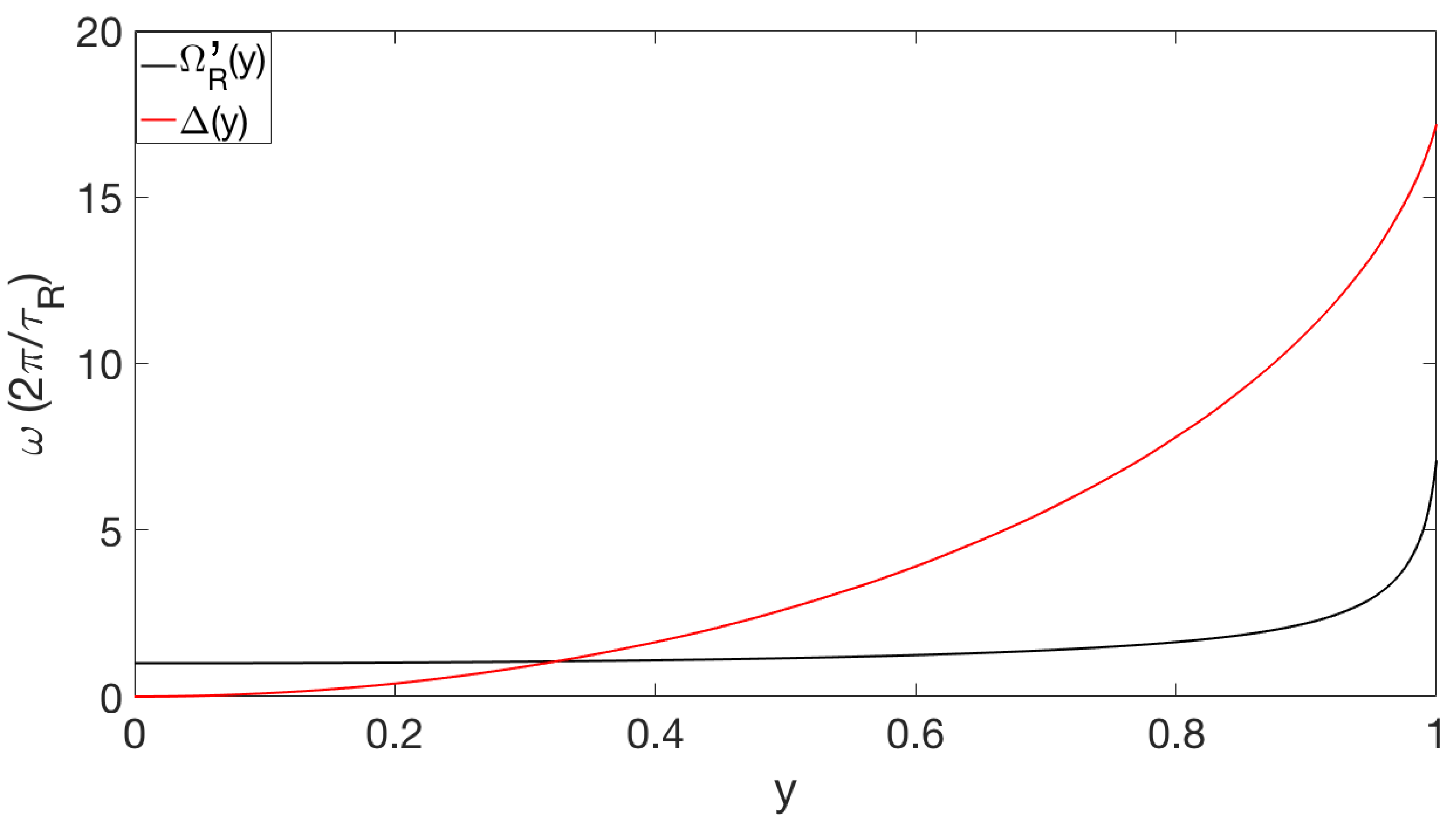}
\caption{\small{Variation of the comoving Rabi frequency $\Omega_R'$ (black) and the detuning $\Delta$ (red) with position $y$ due to the relativistic flow of atoms for the solution depicted in figure \ref{fig: cartflow} above.}}  
\label{fig: freqvar}
\end{figure}
given a laboratory frame optical frequency $\omega_E$ and an atomic transition frequency $\omega_0$. Note the time derivative indicated by dots is with respect to $t'$ as defined above and not $t$. Hence we solve the Bloch equations in the comoving frame of atoms at each position $y$; transforming the results back to the laboratory frame simply means reversing the Lorentz transform in eqs. \ref{simpleLorentz} to express them in terms of laboratory time $t$. The flow induces a $y$-dependent time dilation for the atoms, $t' = t / \gamma(y)$. Additionally the detuning increases with $y$, reducing the medium's response away from the origin. The spontaneous emission rate $\Gamma$ is independent of $y$ because the Bloch equations are being solved in frames comoving with the atoms; in these frames the atoms perceive the same vacuum as they would if they were at rest\footnote{This applies for atoms undergoing inertial motion; the same is not true for accelerated atoms. We will discuss this further in the next section.}.

We solve the Bloch equations \ref{eq: generalBloch} by numerically integration, replacing the time derivative with a central finite difference approximation. Figure \ref{fig: cartflow} shows the physical quantities $W, |P|^2$ associated with the solution $\mathbf{b}(y, t)$ to equations \ref{eq: generalBloch} in the laboratory frame, assuming the linear velocity profile from above with $\Omega = 0.99 c / y_{max}$, no spontaneous emission $\Gamma=0$ and resonant interactions in the laboratory frame $\omega_E=\omega_0$. The atoms are initially prepared in a state $\mathbf{b} = \left(1/\sqrt{2}, -1/\sqrt{2}, 0 \right)$. For larger values of $y \approx 0.8$ the Doppler shift is obvious where inversion and polarisation oscillations are weakened, as the strong detuning suppresses light-matter coupling. Close to the channel's edge $y=1$ time dilation dominates and the system's evolution is retarded as seen by the laboratory observer, while the diverging Rabi frequency boosts the inversion. Clearly the result at $y=0$ corresponds to the stationary case. There is however an intermediate region around $y=0.25$ where the inversion oscillations are strengthened while the polarisation is periodically driven to zero; here the Rabi frequency is close to its stationary value while the Doppler detuning $\Delta = \omega_0 - \omega_E'$ is small relative to $\omega_0$ (see figure \ref{fig: freqvar}).  

\begin{figure}[ht] 
\centering
\includegraphics[width=0.9\linewidth]{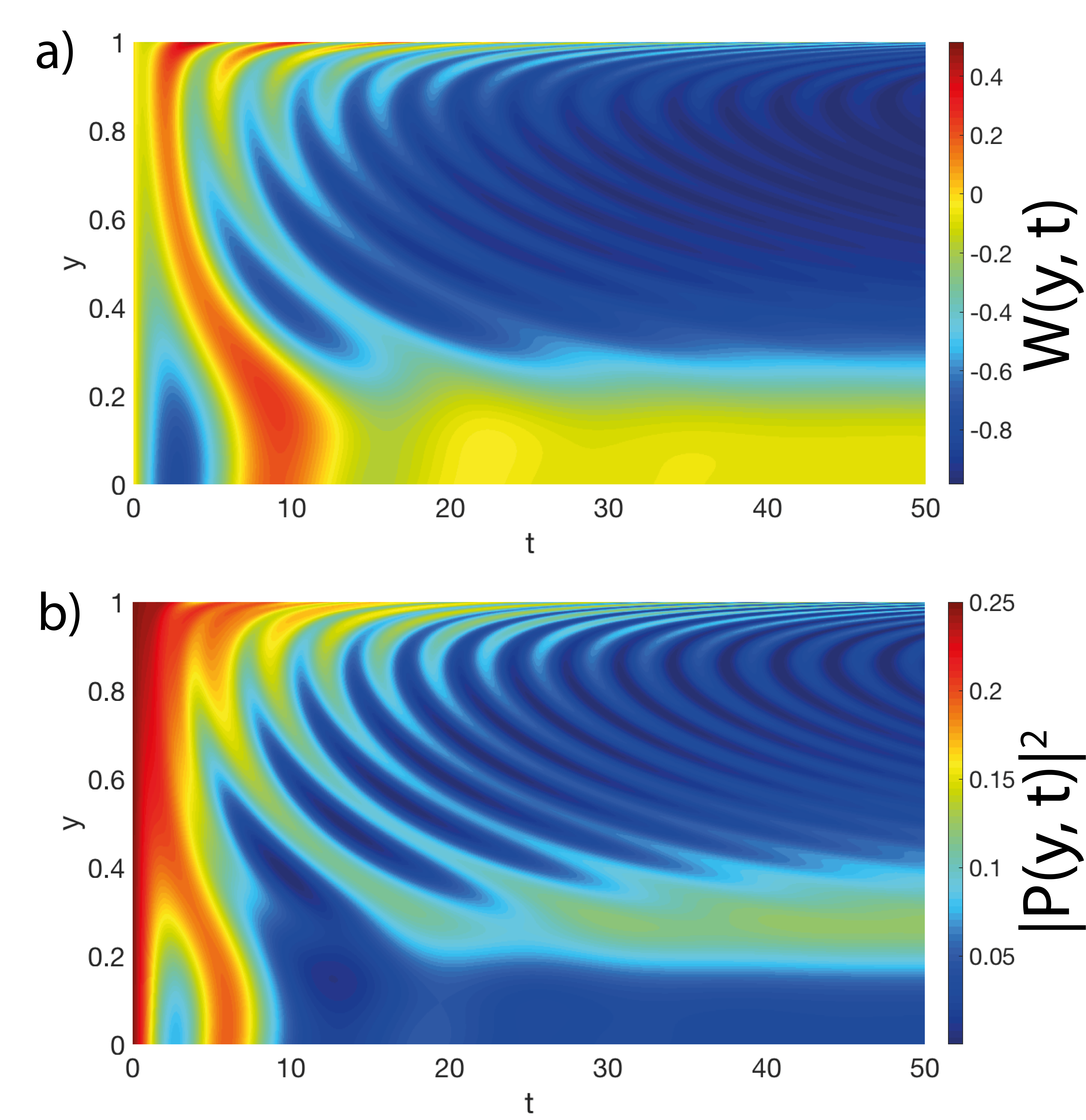}
\caption{\small{Inversion $W$ \textbf{a)} and polarisation magnitude $|P|^2$ \textbf{b)} as defined by equations \ref{eq: W} and \ref{eq: P} respectively, induced by a coherent, homogeneous source in a line of atoms flowing relative to both the source and observer with velocity profile $\mathbf{v}(y) = \Omega y \hat{\mathbf{x}}$. The only difference from the scenario depicted in figure \ref{fig: cartflow} is that here the atoms are allowed to spontaneously decay at a rate $\Gamma = \omega_0 / 50$. This leads to decoherence of the optical response and oscillations in the atomic inversion and polarisation decay over time.  One time unit corresponds to one sixth of the stationary Rabi oscillation period $\tau_R$.}}  
\label{fig: decoheringcartflow}
\end{figure}

Repeating the simulation with spontaneous emission $\Gamma = \omega_0 / 50$ yields figure \ref{fig: decoheringcartflow}. As expected the spontaneous process leads to decoherence of the light-matter interaction overall, but it is reduced in the same region where figure \ref{fig: cartflow} shows enhanced coupling. For the parameters chosen, both the detuning and $\Omega_R$ are small relative to $\omega_0$ so the rotating wave approximation may be valid to some extent. In this case, steady state solutions to the Bloch equations are easily obtained by setting the time derivatives to $0$. Given the Doppler detuning , the steady state values of the inversion and polarisation magnitude are respectively \cite{HumboltNotes}

\begin{equation} \label{eq: steadystatesols}
\begin{split}
&W_{ss}(y) = \frac{ 2 \Omega_R'^2}{\Gamma^2 + 4 \Delta^2 + 2 \Omega_R'^2} - 1\\
&|P|_{ss}(y) =  \frac{ \Omega_R' \sqrt{\Gamma^2 + 4 \Delta^2}}{\Gamma^2 + 4 \Delta^2 + 2 \Omega_R'^2}
\end{split}
\end{equation}   

\begin{figure}[ht] 
\centering
\includegraphics[width=0.8\linewidth]{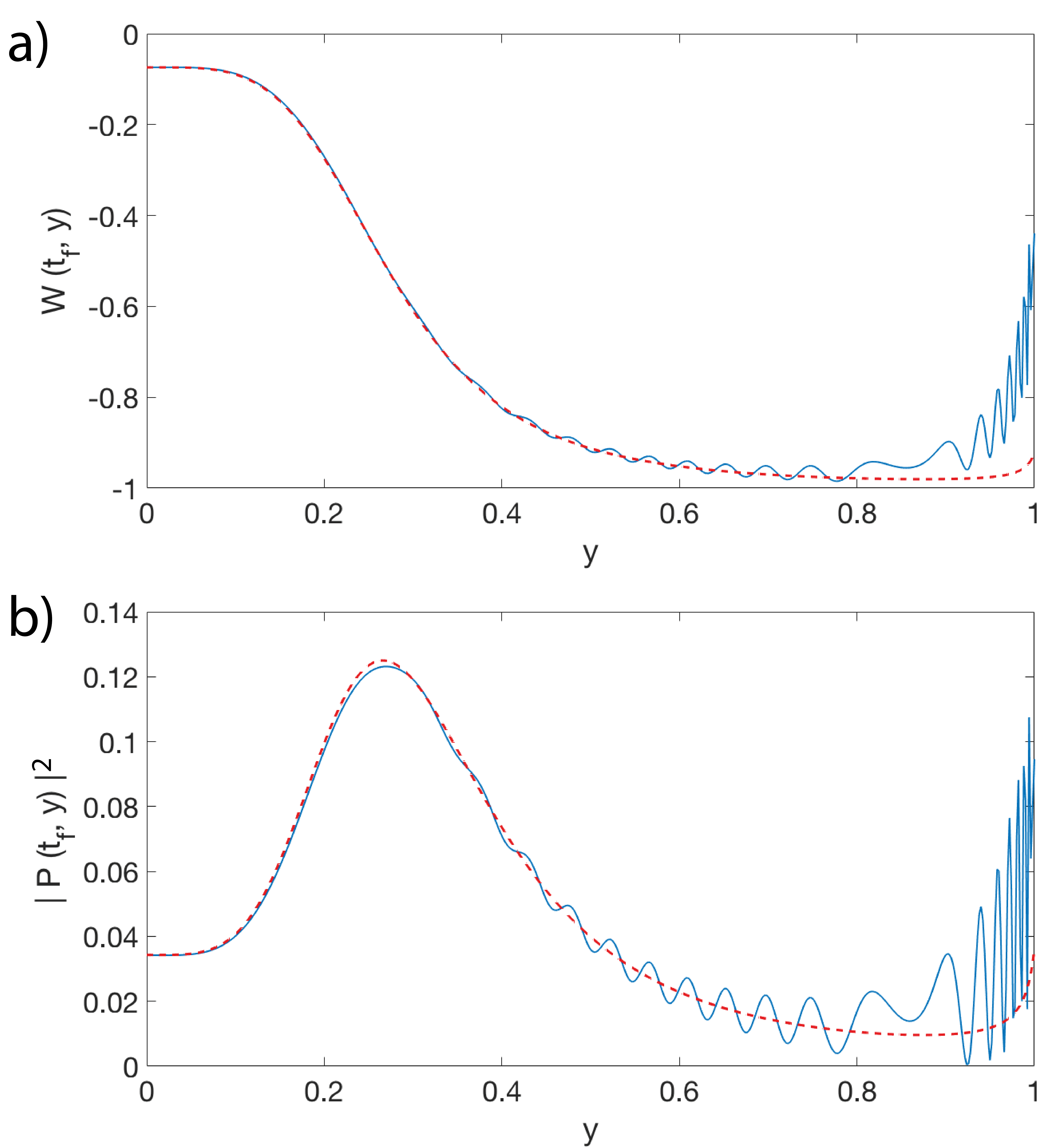}
\caption{\small{Lineout of the inversion \textbf{a)} and polarisation magnitude \textbf{b)} data shown in figure \ref{fig: decoheringcartflow} at the final time $t_f = 50$, solid blue lines. Dashed red lines indicate the steady state solutions to the optical Bloch equations in the rotating wave approximation.}} 
\label{fig: steadystatecomp}
\end{figure}

\begin{figure}[ht] 
\centering
\includegraphics[width=\linewidth]{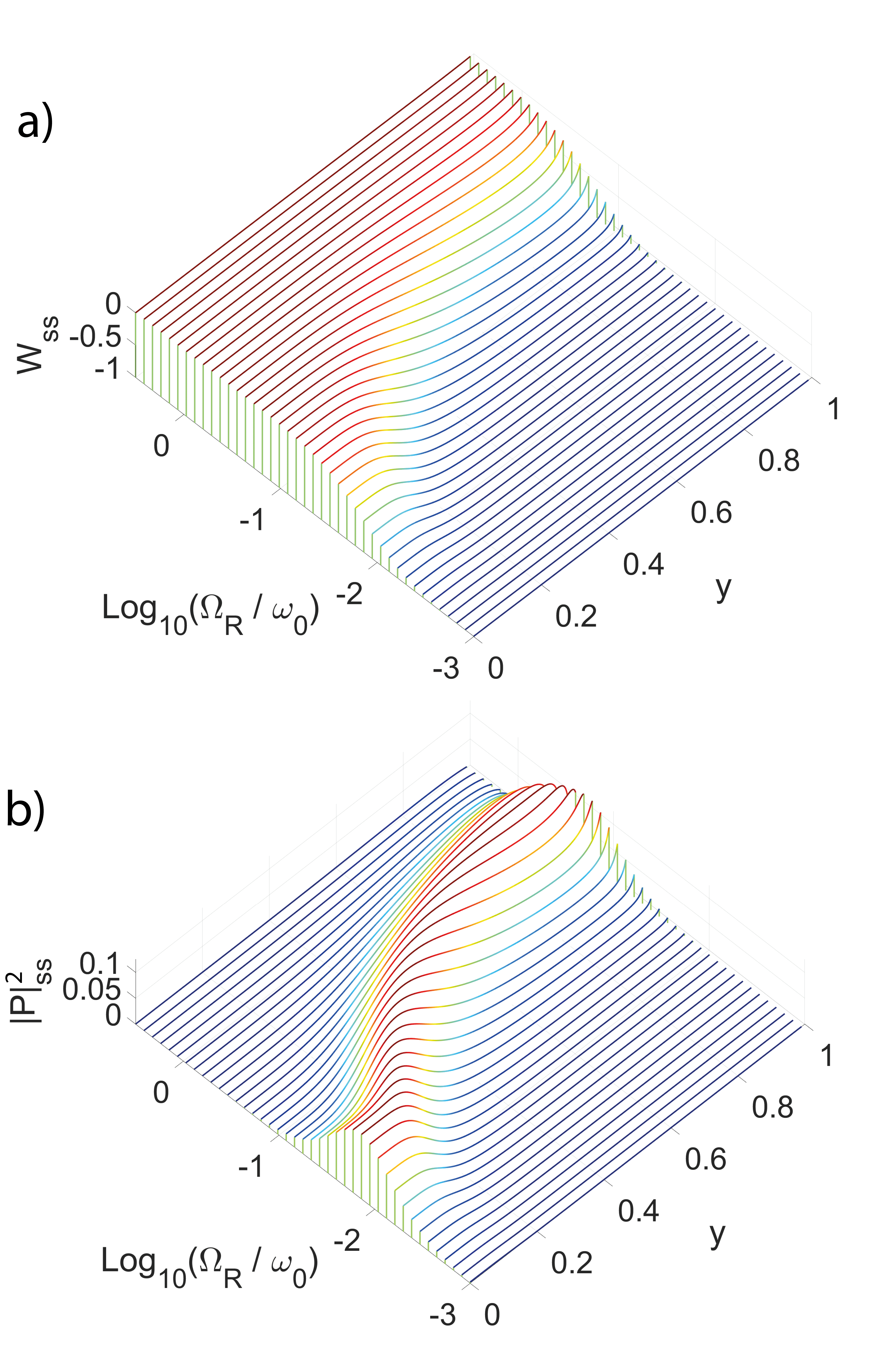}
\caption{\small{Plots of the steady state inversion $W_{ss}$ \textbf{a)} and polarisation $|P|_{ss}^2$ \textbf{b)} based on the analytical solutions in equations \ref{eq: steadystatesols} for varying size of the laboratory frame Rabi frequency $\Omega_R$ relative to the atomic transition frequency $\omega_0$.}} 
\label{fig: steadystatevsRabiplots}
\end{figure}

Figure \ref{fig: steadystatecomp} compares these solutions to the final inversion and polarisation shown in figure \ref{fig: decoheringcartflow}. For $y<0.8$ the agreement is quite strong. Deviations from the steady state as $y \rightarrow 1$ arise as the rotating wave approximation breaks down as both the detuning and Rabi frequency begin to diverge. Figure \ref{fig: freqvar} shows the detuning increases slowly while the Rabi frequency is roughly constant for small $y$; this modest range of detuning causes the polarisation peak around $y=0.3$. Where the detuning is weak, the atomic ensemble approaches the threshold of population inversion ($W_{ss}=0$) at which it becomes transparent, limited by the size of $\Gamma$. This transparency is suppressed by the detuning at larger values of $y$, as light-matter coupling becomes less efficient and larger fractions of the population remain in the ground state. Increasing the laboratory frame Rabi frequency up to $\omega_0$ causes the steady state transparency region to broaden and the polarisation peak to shift outwards to larger $y$. In the strong pumping regime the entire channel becomes transparent as the gas sits just below the population inversion threshold for all $y$. The steady state inversion and polarisation solutions for different values of the laboratory frame Rabi frequency (\textit{i.e.} $\Omega_R = \Omega_R'(y=0)$) are plotted in figure \ref{fig: steadystatevsRabiplots}.  

\section{Solid Rotation Problem}

Now consider an equivalent system in polar coordinates $(t, r, \theta)$; a rotational atomic flow $v_\theta = \Omega r \mathbf{\hat{\theta}}$ confined to a disc interacting with light with laboratory frame polarisation $\mathbf{E} = E_0 \mathbf{\hat{r}}$, $\mathbf{B} = (E_0/c) \mathbf{\hat{\theta}}$, as might be found in a coaxial cable TEM mode. At the disc's edge, the atoms' tangential velocity reaches $0.99 c$. Again we assume no interactions with the disc's boundary or between the atoms. The frames comoving with the atoms are accelerated and non-inertial, so are not strictly connected to the laboratory frame by Lorentz transforms, which are only defined between pairs of inertial frames. However, the experimental findings of the Wilson-Wilson experiment \cite{Wilson1913, Hertzberg2001} and the concurring results of Canovan and Tucker \cite{Canovan2010} suggest that Lorentz transforming to a local inertial frame, momentarily-comoving with the rotating atoms, yields at least an accurate approximation for the rotating frame electromagnetic fields.
Moreover, the work of Gr{\o}n \cite{Gron1975, Gron1977, Gron2007} highlights that the global geometry of the rotating disc is non-Euclidean and described by a Galilean transformation relative to the laboratory frame; however rotating observers will locally see relativistic effects, for example Lorentz length contraction and time dilation, on observables in their proper frame compared to laboratory measures of the same. Time dilation is generally accepted as necessary when describing the trajectory of a relativistically rotating object in laboratory coordiantes \cite{Davies1996, Bell1987, Jin2014}. As such we expect the external electromagnetic field seen by the atoms to be equivalent to that seen by an instantaneously comoving Lorentz observer.  We therefore define a series of local frames, each at a fixed radius $r=R$, by 
\begin{equation} \label{eq: pseudoLorentz}
\begin{split}
t' &= \gamma(R) \left( t - \Omega R^2 / c^2 \theta \right)  \\
\theta' &= \gamma(R) \left( \theta - \Omega t \right)\\ 
R' &= R
\end{split}
\end{equation} 
where $\gamma(R) = 1 / \sqrt{ 1 - {\left( \Omega R / c \right)}^2 }$. We emphasise that this does not define the global geometry of the rotating disc, but correctly relates local comoving observables to their laboratory frame equivalents. The fields seen by the atoms in this frame are then obtained by the same transforms (eqs. \ref{eq:SRfields}) as used in the Cartesian case:
\begin{equation} \label{eq: rotatingfields}
\begin{split}
\mathbf{E}' &= \gamma(R) E_0 \mathbf{\hat{r}} \\
\mathbf{B}' &= \frac{E_0}{c} \mathbf{\hat{\theta}} - \frac{\Omega R}{c^2} E_0 \mathbf{\hat{z}}
\end{split}
\end{equation}
Within these approximations, we expect the results from the linear flow problem discussed in the previous section to translate over to this rotational case. There is however one major distinction from the inertial case, in that the atoms experience a thermal vacuum as they are rotationally accelerated. This is a variant on the Unruh effect \cite{Unruh1976}, originally derived for linear acceleration. The modified vacuum modes allow atoms in the ground state to be spontaneously excited, as well modify the spontaneous emission rate from its inertial value $\Gamma$. Several authors have explored these two effects on rotating two-level systems \cite{Audretsch1995}, in particular Jin \textit{et. al.} \cite{Jin2014} for a two-level atom in the prescence of a fluctuating electromagnetic vacuum field. Using their results, the new spontaneous emission rate is
\begin{equation} \label{eq: spontemm}
\begin{split}
\Gamma_{\downarrow}(R) = &\Gamma \left[1 +  \frac{a^2}{c^2 {\omega_0}^2} \right.\\
 &+ \left. \left( \frac{a^2}{8 c^2 {\omega_0}^2} + \frac{5 a^3}{16 \sqrt{3} c^3 {\omega_0}^3}\right)e^{-2\sqrt{3} \frac{\omega_0 c}{a}} \right]
\end{split}
\end{equation}
while the spontaneous excitation rate is
\begin{equation} \label{eq: spontex}
\Gamma_{\uparrow}(R) = \Gamma \left( \frac{a^2}{8 c^2 {\omega_0}^2} + \frac{5 a^3}{16 \sqrt{3} c^3 {\omega_0}^3}\right)e^{-2 \sqrt{3} \frac{\omega_0 c}{a}}
\end{equation}
given that the atoms' proper acceleration is $a = \gamma^2 \Omega^2 R$. These modify the Bloch equations slightly;
\begin{equation} \label{eq: generalaccBloch}
\begin{split}
\dot{b_1} &= - \omega_0 b_2 + \Omega_R' \sin{(\omega_E' t')}b_3  - \frac{\Gamma_\downarrow + \Gamma_\uparrow}{2} b_1\\
\dot{b_2} &= \omega_0 b_1 - \Omega_R' \cos{(\omega_E' t')} b_3  - \frac{\Gamma_\downarrow + \Gamma_\uparrow}{2} b_2\\
\dot{b_3} &= -\Omega_R' \sin{(\omega_E' t')}b_1 + \Omega_R' \cos{(\omega_E' t')} b_2\\
 &-\left( \Gamma_\downarrow + \Gamma_\uparrow \right) b_3 + \Gamma_\uparrow - \Gamma_\downarrow 
\end{split}
\end{equation} 
and hence their steady state solutions differ from eq. \eqref{eq: steadystatesols}; 
\begin{equation} \label{eq: accsteadystatesols}
\begin{split}
&W_{ss}(R) = \frac{ \left(\Gamma_{\uparrow} - \Gamma_{\downarrow}\right) \left( {\left(\Gamma_{\uparrow} + \Gamma_{\downarrow}\right)}^2 + 4 \Delta^2 \right)}{\left(\Gamma_{\uparrow} + \Gamma_{\downarrow}\right){\left(\Gamma_{\uparrow} + \Gamma_{\downarrow}\right)}^2 + 4 \Delta^2 + 2 \Omega_R'^2}\\
&|P|_{ss}(R) =  \frac{|\Gamma_{\uparrow} - \Gamma_{\downarrow}| \Omega_R' \sqrt{{\left(\Gamma_{\uparrow} + \Gamma_{\downarrow}\right)}^2 + 4 \Delta^2}}{\left(\Gamma_{\uparrow} + \Gamma_{\downarrow}\right){\left(\Gamma_{\uparrow} + \Gamma_{\downarrow}\right)}^2 + 4 \Delta^2 + 2 \Omega_R'^2}
\end{split}
\end{equation}   
\begin{figure}[hb] 
\centering
\includegraphics[width=\linewidth]{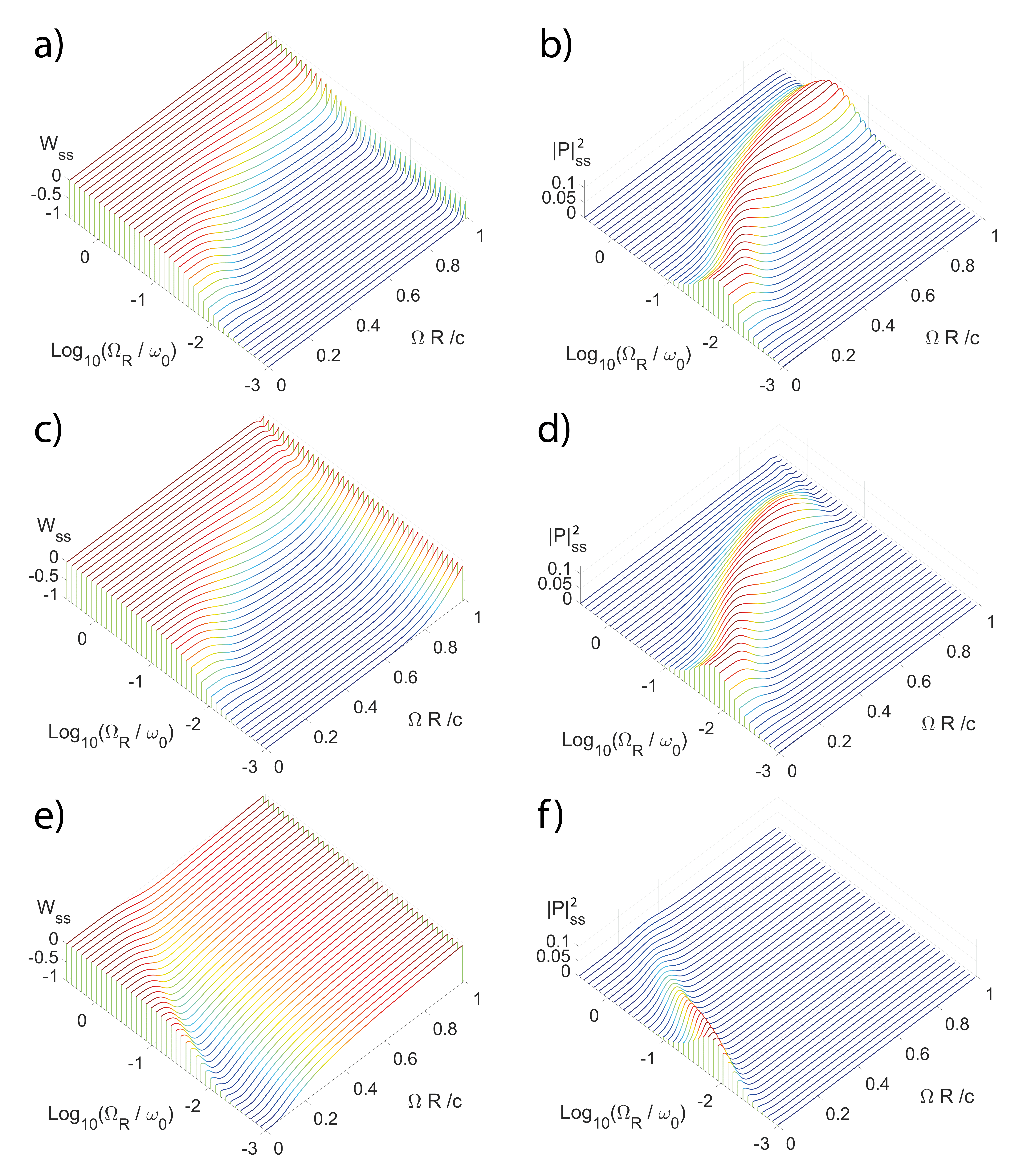}
\caption{\small{Plots of the steady state inversion $W_{ss}$ \textbf{a), c), e)} and polarisation magnitude $|P|^2_{ss}$ \textbf{b), d), f)} based on the analytical solutions in equations \eqref{eq: accsteadystatesols} for varying size of the laboratory frame Rabi frequency relative to the atomic transition frequency. The angular frequency of the disc differs in each row of subfigures to demonstrate acceleration effects: $\Omega = 0.099 \omega_0$ for \textbf{a), b)}, $\Omega = 0.99 \omega_0$ for \textbf{c), d)}, $\Omega = 19.8 \omega_0$ for \textbf{e), f)} }} 
\label{fig: accsteadystatevsRabiplots}
\end{figure}
Clearly the form of the centripetal acceleration means it will be more significant for small radius discs with high angular frequencies than wide, slowly rotating ones. In either case, we intuit that in the extreme limit $\Omega R \rightarrow c$ both the steady state inversion and polarisation tend to zero. This is verified by plotting their solutions as per eq. \eqref{eq: accsteadystatesols}, for increasing rotation frequencies while reducing the maximum radius of the disk to keep $\Omega R = 0.99c$ at the disc's edge constant (figure \ref{fig: accsteadystatevsRabiplots}). The acceleration defines an effective temperature for the thermal vacuum, which diverges with $\gamma^2$, as compared to the driving field which diverges with $\gamma$. As pointed out by previous authors, this introduces a new component to the Lamb shift between atomic energy levels, in addition to its inertial value \cite{Audretsch1995, Jin2014}. Taking the inertial Lamb shift to be negligible, for the parameters we use here ($\Gamma << \omega_0$) we find the additional shift due to rotation is less than $3 \%$ of $\omega_0$ at the disc's edge in the extreme case where $\Omega \approx 10^5 \omega_0$, hence we ignore it. Strong acceleration prevents significant polarisation developing at larger radii, while also boosting inversion to an upper limit of $W=0$ (see $\Omega R > 0.9c$ in plots \ref{fig: accsteadystatevsRabiplots} \textbf{c)}, \textbf{d)} and $\Omega R > 0.4c$ in \ref{fig: accsteadystatevsRabiplots} \textbf{c)}, \textbf{d)}). Both are understandable consequences of the atoms experiencing an effective thermal environment due to their non-inertial motion. Coherence is lost as the bath temperature diverges with relativistic acceleration and the population is driven to the threshold of inversion. The only other significant difference is that the spatial coordinate in the direction of motion now imposes periodic boundary conditions on the external fields and the atoms' response, in both laboratory and rotating frames. 

\begin{figure} 
\includegraphics[width=\linewidth]{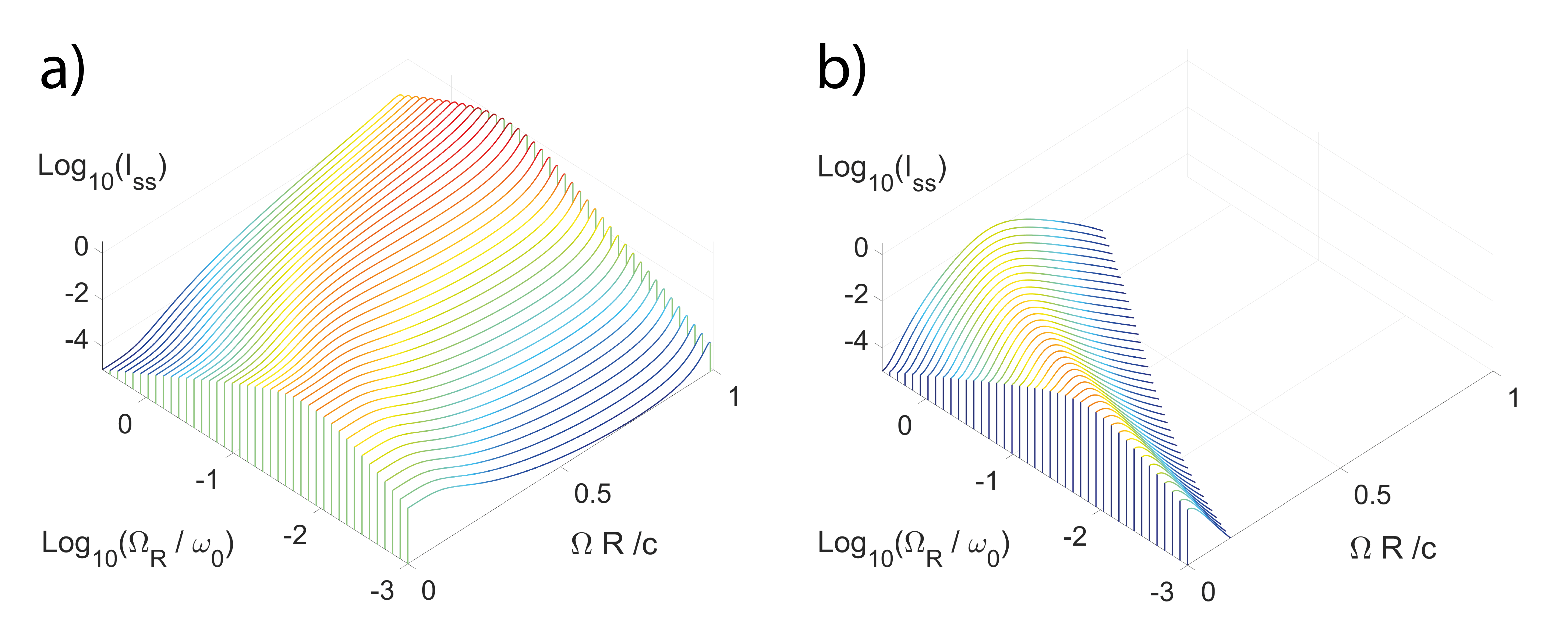}
\caption{\small{Steady-state intensity $I_{ss}$ (viewed on a logarithmic scale relative to ${\epsilon_0}^3 c {p_0}^2/ 2$) of the stimulated emission of the rotating two-level atomic gas, under weak \textbf{a)} and strong \textbf{b)} acceleration, due to the polarisation distributions shown in figure \ref{fig: accsteadystatevsRabiplots} \textbf{b)} and \textbf{f)} respectively.  }}  
\label{fig: stimIntVis}
\end{figure}

Lastly we calculate and visualise the intensity of stimulated emission seen in the laboratory frame due to the steady-state atomic polarisation solutions shown in figure \ref{fig: accsteadystatevsRabiplots} \textbf{b),  f))}. The permittivity of noble gases whose interaction with light can be well described by a two-level system is close to that of vacuum, so in the co--rotating atomic frame the electric field is $\mathbf{E}'_{ss} = \epsilon_0 p_0 |P(r)|_{ss} \mathbf{\hat{r}}$ with the orientation fixed by the pump field, $p_0$ the atomic dipole density and $\epsilon_0$ the permittivity of vacuum. This time-varying electric field will have an associated magnetic field $\mathbf{B}'_{ss} = \left( \epsilon_0 p_0 |P(r)|_{ss} / c\right) \hat{\mathbf{\theta}}$. Using the inverse of the transformations in equations \ref{eq:SRfields} gives the electric and magnetic fields in the laboratory due to the gas's polarisation:
\begin{equation} \label{eq: flourescentfields}
\begin{split}
\mathbf{E}_{ss} &= \gamma(r) \epsilon_0 p_0|P(r)|_{ss} \mathbf{\hat{r}} \\
\mathbf{B}_{ss} &= \frac{\epsilon_0 p_0 |P(r)|_{ss}}{c} \left(\mathbf{\hat{\theta}} - \gamma(r) \frac{\Omega r}{c} \mathbf{\hat{z}} \right)\\ 
\end{split}
\end{equation}           
for which the Poynting vector is 
\begin{equation} \label{eq: flourescentPoynting}
\mathbf{S}_{ss} =  {\epsilon_0}^3 c \gamma(r) {p_0}^2 {|P(r)|_{ss}}^2 \left( \gamma(r) \frac{\Omega r}{c} \mathbf{\hat{\theta}} + \mathbf{\hat{z}} \right).
\end{equation}        
The observed intensity will be the root-mean-square magnitude of this:
\begin{equation} \label{eq: flourescentInt}
I_{ss} = \frac{1}{2} {\epsilon_0}^3 c {\gamma(r)}^2 {p_0}^2 {|P(r)|_{ss}}^2.
\end{equation}        
The steady-state intensity seen by a laboratory frame observer as a function of disc radius and driving strength is shown in figure \ref{fig: stimIntVis} for disc rotation frequencies $\Omega = 0.099 \omega_0$ (\textbf{a)} and $\Omega = 19.8 \omega_0$ (\textbf{b)}. When the gas is weakly pumped $\Omega_R \approx 0.01 \omega_0$ the emission is concentrated around the origin. In an intermediate regime $\Omega_R \approx 0.1 \omega_0$ emission peaks around an inner ring (as may be inferred from the polarisation patterns in figure \ref{fig: accsteadystatevsRabiplots}) as well the around disc's edge. The edge emission dominates under strong pumping $\Omega_R \approx \omega_0$. This is easily understood as the stimulated intensity is proportional to ${\gamma^2}$, so it will be strongest where the gas's velocity approaches the relativistic limit, provided significant polarisation is supported by pumping there. However this is also where atomic coherence can be destroyed by the acceleration-induced thermal vacuum, hence stimulated emission is suppressed in rapidly rotating discs, particularly at outer radii where $a(R)$ diverges.

\section{Discussion}
The final inversion and polarisation lineouts shown in figure \ref{fig: steadystatecomp} \textbf{a)} and \textbf{b)} respectively allow two observations to be made from the laboratory observer's perspective. Firstly, in a modestly-pumped regime Doppler detuning at larger radii leads to only a narrow region of partially excited gas (where $W > -1$) around $r=0$. Laboratory observers perceive a transparent tunnel through the rotating gas, surrounded by absorbing atoms in their ground states. Secondly, the ring of boosted polarisation implies that observers should see a doughnut-like intensity structure in the emitted fluorescence, despite the gas being pumped homogeneously in their reference frame. This appears true when the gas is modestly pumped ($\Omega_R \approx 0.1 \omega_0$) based on the results of figure \ref{fig: stimIntVis}. With weak pumping the emission consists of a central spot, while under strong pumping it is most intense around the disc's edge where relativistic enhancement effects are greatest. Further from figure \ref{fig: steadystatevsRabiplots}\textbf{a)} the central transparency region widens as pumping strength increases. 

We note that analysing the light-matter interaction in the atomic rest frame greatly simplifies the problem. In the laboratory frame, the atoms acquire significant magnetic moments due to their relativistic motion \cite{Hnizdo2012} so their interaction with both the electric and magnetic fields would have to be accounted for. Even to first order in $v/c$, the atomic Hamiltonian would be modified by a R\"{o}ntgen term $\propto \mathbf{v} \cdot \left( \mathbf{B} \times \mathbf{\mu} \right)$, which would shift the energy separation between ground and excited states depending on $r$ \cite{Sonnleitner2017}.  
  
Throughout this work we have assumed no interactions between atoms in the gas, so our result is effectively many iterations of solutions to the Bloch equations for a single-atom at different radii. This assumption should be valid for noble gases, whose atoms fluoresce but interact only weakly through Van der Walls forces, which should be negligible if the gas is sufficiently dilute.
  
Part of the novelty of our work is its inherently quantum nature; existing models for electromagnetism in moving media (such as the Gordon metric) are strictly only applicable to classical, continuous media and are not suitable for our system. Many quantum optical processes, such as the pumping of two-level atoms considered here, fall beyond the scope of Maxwell's equations and do not have manifestly covariant models. Hence describing their physics in relativistic, particularly non-inertial, frames is non-trivial and a worthwhile consideration.
  
Using instantaneously comoving inertial frames to describe the rotating system neglects the nonlocality discussed by Mashhoon \cite{Mashhoon2011, Mashhoon2012} which would require the past field's history, as well as a description of the disc being accelerated from rest to the uniform rate $\Omega$, to be accounted for. If the disc is already rotating with angular frequency $\Omega$ when the light source is switched on in the laboratory frame, we may assume this nonlocality has a small, transient effect which is negligible at longer times. % The Unruh effect may also be significant here; the atoms experience a constant centripetal acceleration, so from an extension Unruh's original analysis \cite{Unruh1976} they see a non-trivial vacuum \cite{Bell1987}. A potential consequence would be that atoms may be excited by Unruh radiation, leading to an extra term in eq. \ref{eq: Liouville} describing acceleration-dependent spontaneous absorbtion. The Unruh effect for detectors with discrete energy levels in uniform circular motion around a cavity has been explored by Levin, Peleg and Pereg \cite{Levin1993} and Davies, Dray and Manogue in free space \cite{Davies1996}. Both parties conclude that the detector remains unexcited providing its rotational velocity is subluminal, \textit{i.e.} $\Omega R < c$. The Unruh effect was analysed more generally by Korsbakken and Leinaas \cite{Korsbakken2004} for systems undergoing both linear and rotational acceleration. They found that detector excitation is possible with an additional, weaker linear acceleration from negative energy modes (the same which give rise to amplification in the Penrose process \cite{Penrose1969, Penrose1971} and the Zel'dovich effect \cite{zeldovich}), but again only when the detector rotates superluminally. Hence we take the Unruh effect to be irrelevant for the two-level atoms considered here.

If linear polarisation $\mathbf{E} \parallel \mathbf{\hat{y}}$ was used in section 3, the electric field $\mathbf{E}'$ would break rotational symmetry. This would result in the Rabi frequency varying with $\theta$ and a much more complicated system, where the atoms would experience periodic driving. 
\section{Conclusions}
The optical Bloch equations have been solved to predict the response of two-level atoms in relativistic linear and rotational motion with respect to a laboratory frame continuous wave light source and observer. At atomic velocities approaching the speed of light, a combination of  time dilation and relativistic detuning of the optical carrier and Rabi frequencies modify the emission seen from the flowing gas significantly compared to when the gas is stationary in the laboratory frame. When the gas is weakly and homogeneously pumped, a laboratory observer will see a ring structure in the rotating gas's stimulated emission and perceive the centre of rotation as being transparent. The generalised Unruh effect suppresses this emission for atoms undergoing strong centripetal acceleration. These results may be relevant to studies of radiation from rapidly rotating astronomical bodies. Observation of such effects would prove useful in validating the instantaneously-comoving frame approach to relativistic rotation of Einstein and Laub as a practical approximation over more complicated methods.     

{\textit{Acknowledgements}}--- C.M. acknowledges studentship funding from EPSRC under CM-CDT Grant No. EP/L015110/1. M.C. acknowledges support from EPSRC/UKRI (Innovate fellowship EP/S001573/1). F.B. acknowledges support from the German Max Planck Society for the Advancement of Science (MPG), in particular the IMPP partnership between Scottish Universities and MPG.

%\bibliographystyle{unsrtnat}
%\bibliography{RotatingMaxwellBloch}

\begin{thebibliography}{44}%
\makeatletter
\providecommand \@ifxundefined [1]{%
 \@ifx{#1\undefined}
}%
\providecommand \@ifnum [1]{%
 \ifnum #1\expandafter \@firstoftwo
 \else \expandafter \@secondoftwo
 \fi
}%
\providecommand \@ifx [1]{%
 \ifx #1\expandafter \@firstoftwo
 \else \expandafter \@secondoftwo
 \fi
}%
\providecommand \natexlab [1]{#1}%
\providecommand \enquote  [1]{``#1''}%
\providecommand \bibnamefont  [1]{#1}%
\providecommand \bibfnamefont [1]{#1}%
\providecommand \citenamefont [1]{#1}%
\providecommand \href@noop [0]{\@secondoftwo}%
\providecommand \href [0]{\begingroup \@sanitize@url \@href}%
\providecommand \@href[1]{\@@startlink{#1}\@@href}%
\providecommand \@@href[1]{\endgroup#1\@@endlink}%
\providecommand \@sanitize@url [0]{\catcode `\\12\catcode `\$12\catcode
  `\&12\catcode `\#12\catcode `\^12\catcode `\_12\catcode `\%12\relax}%
\providecommand \@@startlink[1]{}%
\providecommand \@@endlink[0]{}%
\providecommand \url  [0]{\begingroup\@sanitize@url \@url }%
\providecommand \@url [1]{\endgroup\@href {#1}{\urlprefix }}%
\providecommand \urlprefix  [0]{URL }%
\providecommand \Eprint [0]{\href }%
\providecommand \doibase [0]{http://dx.doi.org/}%
\providecommand \selectlanguage [0]{\@gobble}%
\providecommand \bibinfo  [0]{\@secondoftwo}%
\providecommand \bibfield  [0]{\@secondoftwo}%
\providecommand \translation [1]{[#1]}%
\providecommand \BibitemOpen [0]{}%
\providecommand \bibitemStop [0]{}%
\providecommand \bibitemNoStop [0]{.\EOS\space}%
\providecommand \EOS [0]{\spacefactor3000\relax}%
\providecommand \BibitemShut  [1]{\csname bibitem#1\endcsname}%
\let\auto@bib@innerbib\@empty
%</preamble>
\bibitem [{\citenamefont {Nikoli\'{c}}(2000)}]{Nikolic2000}%
  \BibitemOpen
  \bibfield  {author} {\bibinfo {author} {\bibfnamefont {H.}~\bibnamefont
  {Nikoli\'{c}}},\ }\href {\doibase 10.1103/PhysRevA.61.032109} {\bibfield
  {journal} {\bibinfo  {journal} {Phys. Rev. A}\ }\textbf {\bibinfo {volume}
  {61}},\ \bibinfo {pages} {032109} (\bibinfo {year} {2000})}\BibitemShut
  {NoStop}%
\bibitem [{\citenamefont {Rizzi}\ and\ \citenamefont
  {Ruggiero}(2004)}]{Rizzi2004}%
  \BibitemOpen
  \bibfield  {author} {\bibinfo {author} {\bibfnamefont {G.}~\bibnamefont
  {Rizzi}}\ and\ \bibinfo {author} {\bibfnamefont {M.~L.}\ \bibnamefont
  {Ruggiero}},\ }\href@noop {} {\emph {\bibinfo {title} {``{Relativity in
  Rotating Frames }''}}}\ (\bibinfo  {publisher} {Kluwer Academic},\ \bibinfo
  {year} {2004})\BibitemShut {NoStop}%
\bibitem [{\citenamefont {Kassner}(2012)}]{Kassner2012}%
  \BibitemOpen
  \bibfield  {author} {\bibinfo {author} {\bibfnamefont {K.}~\bibnamefont
  {Kassner}},\ }\href {\doibase 10.1119/1.4730925} {\bibfield  {journal}
  {\bibinfo  {journal} {Am. J. Phys.}\ }\textbf {\bibinfo {volume} {80}},\
  \bibinfo {pages} {772} (\bibinfo {year} {2012})}\BibitemShut {NoStop}%
\bibitem [{\citenamefont {Ekşİ}\ and\ \citenamefont
  {Alpar}(2005)}]{Eksi2005}%
  \BibitemOpen
  \bibfield  {author} {\bibinfo {author} {\bibfnamefont {K.~Y.}\ \bibnamefont
  {Ekşİ}}\ and\ \bibinfo {author} {\bibfnamefont {M.~A.}\ \bibnamefont
  {Alpar}},\ }\href {\doibase 10.1086/425959} {\bibfield  {journal} {\bibinfo
  {journal} {Astrophys. J.}\ }\textbf {\bibinfo {volume} {620}},\ \bibinfo
  {pages} {390} (\bibinfo {year} {2005})}\BibitemShut {NoStop}%
\bibitem [{\citenamefont {Franklin}(1922)}]{Franklin1922}%
  \BibitemOpen
  \bibfield  {author} {\bibinfo {author} {\bibfnamefont {P.}~\bibnamefont
  {Franklin}},\ }\href {\doibase 10.1073/pnas.8.9.265} {\bibfield  {journal}
  {\bibinfo  {journal} {Proc. Nat. Acad. Sci.}\ }\textbf {\bibinfo {volume}
  {8}},\ \bibinfo {pages} {265} (\bibinfo {year} {1922})}\BibitemShut {NoStop}%
\bibitem [{\citenamefont {Takeno}(1952)}]{Takeno1952}%
  \BibitemOpen
  \bibfield  {author} {\bibinfo {author} {\bibfnamefont {H.}~\bibnamefont
  {Takeno}},\ }\href {\doibase 10.1143/ptp/7.4.367} {\bibfield  {journal}
  {\bibinfo  {journal} {Prog. Theor. Phys.}\ }\textbf {\bibinfo {volume} {7}},\
  \bibinfo {pages} {367} (\bibinfo {year} {1952})}\BibitemShut {NoStop}%
\bibitem [{\citenamefont {Post}(1967)}]{Post1967}%
  \BibitemOpen
  \bibfield  {author} {\bibinfo {author} {\bibfnamefont {E.~J.}\ \bibnamefont
  {Post}},\ }\href {\doibase 10.1103/RevModPhys.39.475} {\bibfield  {journal}
  {\bibinfo  {journal} {Rev. Mod. Phys.}\ }\textbf {\bibinfo {volume} {39}},\
  \bibinfo {pages} {475} (\bibinfo {year} {1967})}\BibitemShut {NoStop}%
\bibitem [{\citenamefont {Strauss}(1974)}]{Strauss1974}%
  \BibitemOpen
  \bibfield  {author} {\bibinfo {author} {\bibfnamefont {M.}~\bibnamefont
  {Strauss}},\ }\href {\doibase 10.1007/BF01811037} {\bibfield  {journal}
  {\bibinfo  {journal} {Int. J. Theor. Phys.}\ }\textbf {\bibinfo {volume}
  {11}},\ \bibinfo {pages} {107} (\bibinfo {year} {1974})}\BibitemShut
  {NoStop}%
\bibitem [{\citenamefont {Nouri-Zonoz}\ \emph {et~al.}(2014)\citenamefont
  {Nouri-Zonoz}, \citenamefont {Ramezani-Aval},\ and\ \citenamefont
  {Gharechahi}}]{NouriZonoz2014}%
  \BibitemOpen
  \bibfield  {author} {\bibinfo {author} {\bibfnamefont {M.}~\bibnamefont
  {Nouri-Zonoz}}, \bibinfo {author} {\bibfnamefont {H.}~\bibnamefont
  {Ramezani-Aval}}, \ and\ \bibinfo {author} {\bibfnamefont {R.}~\bibnamefont
  {Gharechahi}},\ }\href {\doibase 10.1140/epjc/s10052-014-3098-6} {\bibfield
  {journal} {\bibinfo  {journal} {Eur. Phys. J. C}\ }\textbf {\bibinfo {volume}
  {74}},\ \bibinfo {pages} {3098} (\bibinfo {year} {2014})}\BibitemShut
  {NoStop}%
\bibitem [{\citenamefont {Anderson}\ \emph {et~al.}(1994)\citenamefont
  {Anderson}, \citenamefont {Bilger},\ and\ \citenamefont
  {Stedman}}]{Anderson1994}%
  \BibitemOpen
  \bibfield  {author} {\bibinfo {author} {\bibfnamefont {R.}~\bibnamefont
  {Anderson}}, \bibinfo {author} {\bibfnamefont {H.~R.}\ \bibnamefont
  {Bilger}}, \ and\ \bibinfo {author} {\bibfnamefont {G.~E.}\ \bibnamefont
  {Stedman}},\ }\href {\doibase 10.1119/1.17656} {\bibfield  {journal}
  {\bibinfo  {journal} {Am. J. Phys.}\ }\textbf {\bibinfo {volume} {62}},\
  \bibinfo {pages} {975} (\bibinfo {year} {1994})}\BibitemShut {NoStop}%
\bibitem [{\citenamefont {Bolotovskii}\ and\ \citenamefont
  {Stolyarov}(1975)}]{bolotovskii}%
  \BibitemOpen
  \bibfield  {author} {\bibinfo {author} {\bibfnamefont {B.~M.}\ \bibnamefont
  {Bolotovskii}}\ and\ \bibinfo {author} {\bibfnamefont {S.~N.}\ \bibnamefont
  {Stolyarov}},\ }\href {\doibase 10.1070/PU1975v017n06ABEH004403} {\bibfield
  {journal} {\bibinfo  {journal} {Sov. Phys. Usp.}\ }\textbf {\bibinfo {volume}
  {17}},\ \bibinfo {pages} {875} (\bibinfo {year} {1975})}\BibitemShut
  {NoStop}%
\bibitem [{\citenamefont {Minkowski}(1908)}]{Minkowski1908}%
  \BibitemOpen
  \bibfield  {author} {\bibinfo {author} {\bibfnamefont {H.}~\bibnamefont
  {Minkowski}},\ }\href@noop {} {\enquote {\bibinfo {title} {Nachr. ges.
  wiss},}\ } (\bibinfo {year} {1908})\BibitemShut {NoStop}%
\bibitem [{\citenamefont {Ivezic}(2012)}]{Ivezic2012}%
  \BibitemOpen
  \bibfield  {author} {\bibinfo {author} {\bibfnamefont {T.}~\bibnamefont
  {Ivezic}},\ }\href {\doibase 10.1142/S0217979212500403} {\bibfield  {journal}
  {\bibinfo  {journal} {Int. J. Mod. Phys. B}\ }\textbf {\bibinfo {volume}
  {26}},\ \bibinfo {pages} {1250040} (\bibinfo {year} {2012})}\BibitemShut
  {NoStop}%
\bibitem [{\citenamefont {Tai}(1964)}]{Tai1964}%
  \BibitemOpen
  \bibfield  {author} {\bibinfo {author} {\bibfnamefont {C.}~\bibnamefont
  {Tai}},\ }\href@noop {} {\bibfield  {journal} {\bibinfo  {journal} {Radio
  Science Journal of Research}\ }\textbf {\bibinfo {volume} {69D}},\ \bibinfo
  {pages} {401} (\bibinfo {year} {1964})}\BibitemShut {NoStop}%
\bibitem [{\citenamefont {Tanaka}\ and\ \citenamefont
  {Havana}(1972)}]{Tanaka1972}%
  \BibitemOpen
  \bibfield  {author} {\bibinfo {author} {\bibfnamefont {K.}~\bibnamefont
  {Tanaka}}\ and\ \bibinfo {author} {\bibfnamefont {K.}~\bibnamefont
  {Havana}},\ }\href {\doibase 10.1029/RS007i010p00973} {\bibfield  {journal}
  {\bibinfo  {journal} {Radio Sci.}\ }\textbf {\bibinfo {volume} {7}},\
  \bibinfo {pages} {973} (\bibinfo {year} {1972})}\BibitemShut {NoStop}%
\bibitem [{\citenamefont {Leonhardt}\ and\ \citenamefont
  {Piwnicki}(1999)}]{piwnicki}%
  \BibitemOpen
  \bibfield  {author} {\bibinfo {author} {\bibfnamefont {U.}~\bibnamefont
  {Leonhardt}}\ and\ \bibinfo {author} {\bibfnamefont {P.}~\bibnamefont
  {Piwnicki}},\ }\href {\doibase 10.1103/PhysRevA.60.4301} {\bibfield
  {journal} {\bibinfo  {journal} {Phys. Rev. A}\ }\textbf {\bibinfo {volume}
  {60}},\ \bibinfo {pages} {4301} (\bibinfo {year} {1999})}\BibitemShut
  {NoStop}%
\bibitem [{\citenamefont {Gordon}(1923)}]{Gordon1923}%
  \BibitemOpen
  \bibfield  {author} {\bibinfo {author} {\bibfnamefont {W.}~\bibnamefont
  {Gordon}},\ }\href {\doibase 10.1002/andp.19233772202} {\bibfield  {journal}
  {\bibinfo  {journal} {Ann. Phys. (Leipzig)}\ }\textbf {\bibinfo {volume}
  {377}},\ \bibinfo {pages} {421} (\bibinfo {year} {1923})}\BibitemShut
  {NoStop}%
\bibitem [{\citenamefont {Ehrenfest}(1909)}]{Ehrenfest1909}%
  \BibitemOpen
  \bibfield  {author} {\bibinfo {author} {\bibfnamefont {P.}~\bibnamefont
  {Ehrenfest}},\ }\href {\doibase 10.1007/978-94-017-0528-8_1} {\bibfield
  {journal} {\bibinfo  {journal} {Phys. Z.}\ }\textbf {\bibinfo {volume}
  {10}},\ \bibinfo {pages} {918} (\bibinfo {year} {1909})}\BibitemShut
  {NoStop}%
\bibitem [{\citenamefont {Gr{\o}n}(1975)}]{Gron1975}%
  \BibitemOpen
  \bibfield  {author} {\bibinfo {author} {\bibfnamefont {{\O}.}~\bibnamefont
  {Gr{\o}n}},\ }\href {\doibase 10.1119/1.9969} {\bibfield  {journal} {\bibinfo
   {journal} {Am. J. Phys.}\ }\textbf {\bibinfo {volume} {43}},\ \bibinfo
  {pages} {869} (\bibinfo {year} {1975})}\BibitemShut {NoStop}%
\bibitem [{\citenamefont {Gr{\o}n}\ and\ \citenamefont
  {Hervik}(2007)}]{Gron2007}%
  \BibitemOpen
  \bibfield  {author} {\bibinfo {author} {\bibfnamefont {{\O}.}~\bibnamefont
  {Gr{\o}n}}\ and\ \bibinfo {author} {\bibfnamefont {S.}~\bibnamefont
  {Hervik}},\ }\href@noop {} {\emph {\bibinfo {title} {Einstein's General
  Theory of Relativity: With Modern Applications in Cosmology}}}\ (\bibinfo
  {publisher} {Springer New York},\ \bibinfo {year} {2007})\BibitemShut
  {NoStop}%
\bibitem [{\citenamefont {Wilson}\ and\ \citenamefont
  {Wilson}(1913)}]{Wilson1913}%
  \BibitemOpen
  \bibfield  {author} {\bibinfo {author} {\bibfnamefont {M.}~\bibnamefont
  {Wilson}}\ and\ \bibinfo {author} {\bibfnamefont {H.~A.}\ \bibnamefont
  {Wilson}},\ }\href {\doibase 10.1098/rspa.1913.0067} {\bibfield  {journal}
  {\bibinfo  {journal} {P. Roy. Soc. Lond. A Mat.}\ }\textbf {\bibinfo {volume}
  {89}},\ \bibinfo {pages} {99} (\bibinfo {year} {1913})}\BibitemShut {NoStop}%
\bibitem [{\citenamefont {Hertzberg}\ \emph {et~al.}(2001)\citenamefont
  {Hertzberg}, \citenamefont {Bickman}, \citenamefont {Hummon}, \citenamefont
  {Krause}, \citenamefont {Peck},\ and\ \citenamefont
  {Hunter}}]{Hertzberg2001}%
  \BibitemOpen
  \bibfield  {author} {\bibinfo {author} {\bibfnamefont {J.~B.}\ \bibnamefont
  {Hertzberg}}, \bibinfo {author} {\bibfnamefont {S.~R.}\ \bibnamefont
  {Bickman}}, \bibinfo {author} {\bibfnamefont {M.~T.}\ \bibnamefont {Hummon}},
  \bibinfo {author} {\bibfnamefont {D.}~\bibnamefont {Krause}}, \bibinfo
  {author} {\bibfnamefont {S.~K.}\ \bibnamefont {Peck}}, \ and\ \bibinfo
  {author} {\bibfnamefont {L.~R.}\ \bibnamefont {Hunter}},\ }\href {\doibase
  10.1119/1.1362695} {\bibfield  {journal} {\bibinfo  {journal} {Am. J. Phys.}\
  }\textbf {\bibinfo {volume} {69}},\ \bibinfo {pages} {648} (\bibinfo {year}
  {2001})}\BibitemShut {NoStop}%
\bibitem [{\citenamefont {Einstein}\ and\ \citenamefont
  {Laub}(1908)}]{Einstein1908}%
  \BibitemOpen
  \bibfield  {author} {\bibinfo {author} {\bibfnamefont {A.}~\bibnamefont
  {Einstein}}\ and\ \bibinfo {author} {\bibfnamefont {J.}~\bibnamefont
  {Laub}},\ }\href@noop {} {\bibfield  {journal} {\bibinfo  {journal} {Ann.
  Phys. (Leipzig)}\ }\textbf {\bibinfo {volume} {26}},\ \bibinfo {pages} {532}
  (\bibinfo {year} {1908})}\BibitemShut {NoStop}%
\bibitem [{\citenamefont {Canovan}\ and\ \citenamefont
  {Tucker}(2010)}]{Canovan2010}%
  \BibitemOpen
  \bibfield  {author} {\bibinfo {author} {\bibfnamefont {C.}~\bibnamefont
  {Canovan}}\ and\ \bibinfo {author} {\bibfnamefont {R.}~\bibnamefont
  {Tucker}},\ }\href {\doibase 10.1119/1.3456566} {\bibfield  {journal}
  {\bibinfo  {journal} {Am. J. Phys.}\ }\textbf {\bibinfo {volume} {78}},\
  \bibinfo {pages} {1181} (\bibinfo {year} {2010})}\BibitemShut {NoStop}%
\bibitem [{\citenamefont {Pellegrini}\ and\ \citenamefont
  {Swift}(1995)}]{Pellegrini1995}%
  \BibitemOpen
  \bibfield  {author} {\bibinfo {author} {\bibfnamefont {G.~N.}\ \bibnamefont
  {Pellegrini}}\ and\ \bibinfo {author} {\bibfnamefont {A.~R.}\ \bibnamefont
  {Swift}},\ }\href {\doibase 10.1119/1.17839} {\bibfield  {journal} {\bibinfo
  {journal} {Am. J. Phys.}\ }\textbf {\bibinfo {volume} {63}},\ \bibinfo
  {pages} {694} (\bibinfo {year} {1995})}\BibitemShut {NoStop}%
\bibitem [{\citenamefont {Burrows}(1997)}]{Burrows1997}%
  \BibitemOpen
  \bibfield  {author} {\bibinfo {author} {\bibfnamefont {M.~L.}\ \bibnamefont
  {Burrows}},\ }\href {\doibase 10.1119/1.18689} {\bibfield  {journal}
  {\bibinfo  {journal} {Am. J. Phys.}\ }\textbf {\bibinfo {volume} {65}},\
  \bibinfo {pages} {929} (\bibinfo {year} {1997})}\BibitemShut {NoStop}%
\bibitem [{\citenamefont {Weber}(1997)}]{Weber1997}%
  \BibitemOpen
  \bibfield  {author} {\bibinfo {author} {\bibfnamefont {T.~A.}\ \bibnamefont
  {Weber}},\ }\href {\doibase 10.1119/1.18696} {\bibfield  {journal} {\bibinfo
  {journal} {Am. J. Phys.}\ }\textbf {\bibinfo {volume} {65}},\ \bibinfo
  {pages} {946} (\bibinfo {year} {1997})}\BibitemShut {NoStop}%
\bibitem [{\citenamefont {Ridgely}(1998)}]{Ridgely1998}%
  \BibitemOpen
  \bibfield  {author} {\bibinfo {author} {\bibfnamefont {C.~T.}\ \bibnamefont
  {Ridgely}},\ }\href {\doibase 10.1119/1.18828} {\bibfield  {journal}
  {\bibinfo  {journal} {Am. J. Phys.}\ }\textbf {\bibinfo {volume} {66}},\
  \bibinfo {pages} {114} (\bibinfo {year} {1998})}\BibitemShut {NoStop}%
\bibitem [{\citenamefont {McDonald}(2008)}]{McDonald2008}%
  \BibitemOpen
  \bibfield  {author} {\bibinfo {author} {\bibfnamefont {K.~T.}\ \bibnamefont
  {McDonald}},\ }\href
  {http://physics.princeton.edu/~mcdonald/examples/wilson.pdf} {\enquote
  {\bibinfo {title} {The wilson-wilson experiment},}\ } (\bibinfo {year}
  {2008})\BibitemShut {NoStop}%
\bibitem [{\citenamefont {Mashhoon}(2008)}]{Mashhoon2008}%
  \BibitemOpen
  \bibfield  {author} {\bibinfo {author} {\bibfnamefont {B.}~\bibnamefont
  {Mashhoon}},\ }\href {\doibase 10.1002/andp.200810308} {\bibfield  {journal}
  {\bibinfo  {journal} {Ann. Phys. (Berlin)}\ }\textbf {\bibinfo {volume}
  {17}},\ \bibinfo {pages} {705} (\bibinfo {year} {2008})}\BibitemShut
  {NoStop}%
\bibitem [{\citenamefont {Mashhoon}(2011)}]{Mashhoon2011}%
  \BibitemOpen
  \bibfield  {author} {\bibinfo {author} {\bibfnamefont {B.}~\bibnamefont
  {Mashhoon}},\ }\href {\doibase 10.1002/andp.201010464} {\bibfield  {journal}
  {\bibinfo  {journal} {Ann. Phys. (Berlin)}\ }\textbf {\bibinfo {volume}
  {523}},\ \bibinfo {pages} {226} (\bibinfo {year} {2011})}\BibitemShut
  {NoStop}%
\bibitem [{\citenamefont {Mashhoon}(2012{\natexlab{a}})}]{Mashhoon2012}%
  \BibitemOpen
  \bibfield  {author} {\bibinfo {author} {\bibfnamefont {B.}~\bibnamefont
  {Mashhoon}},\ }\href {\doibase 10.1002/andp.201200208} {\bibfield  {journal}
  {\bibinfo  {journal} {Ann. Phys. (Berlin)}\ }\textbf {\bibinfo {volume}
  {525}},\ \bibinfo {pages} {235} (\bibinfo {year}
  {2012}{\natexlab{a}})}\BibitemShut {NoStop}%
\bibitem [{\citenamefont {Mashhoon}(2012{\natexlab{b}})}]{Mashhoon2012b}%
  \BibitemOpen
  \bibfield  {author} {\bibinfo {author} {\bibfnamefont {B.}~\bibnamefont
  {Mashhoon}},\ }\href@noop {} {\enquote {\bibinfo {title} {{Nonlocal Special
  Relativity: Amplitude Shift in Spin-Rotation Coupling}},}\ } (\bibinfo {year}
  {2012}{\natexlab{b}}),\ \Eprint {http://arxiv.org/abs/1204.6069}
  {arXiv:1204.6069 [gr-qc]} \BibitemShut {NoStop}%
%%CITATION = ARXIV:1204.6069;%%
\bibitem [{\citenamefont {Desloge}\ and\ \citenamefont
  {Philpott}(1987)}]{Desloge1987}%
  \BibitemOpen
  \bibfield  {author} {\bibinfo {author} {\bibfnamefont {E.~A.}\ \bibnamefont
  {Desloge}}\ and\ \bibinfo {author} {\bibfnamefont {R.~J.}\ \bibnamefont
  {Philpott}},\ }\href {\doibase 10.1119/1.15197} {\bibfield  {journal}
  {\bibinfo  {journal} {Am. J. Phys.}\ }\textbf {\bibinfo {volume} {55}},\
  \bibinfo {pages} {252} (\bibinfo {year} {1987})}\BibitemShut {NoStop}%
\bibitem [{\citenamefont {Audretsch}\ \emph {et~al.}(1995)\citenamefont
  {Audretsch}, \citenamefont {M{\"u}ller},\ and\ \citenamefont
  {Holzmann}}]{Audretsch1995}%
  \BibitemOpen
  \bibfield  {author} {\bibinfo {author} {\bibfnamefont {J.}~\bibnamefont
  {Audretsch}}, \bibinfo {author} {\bibfnamefont {R.}~\bibnamefont
  {M{\"u}ller}}, \ and\ \bibinfo {author} {\bibfnamefont {M.}~\bibnamefont
  {Holzmann}},\ }\href {\doibase 10.1088/0264-9381/12/12/010} {\bibfield
  {journal} {\bibinfo  {journal} {Classical Quant. Grav.}\ }\textbf {\bibinfo
  {volume} {12}},\ \bibinfo {pages} {2927} (\bibinfo {year}
  {1995})}\BibitemShut {NoStop}%
\bibitem [{\citenamefont {Jones}(2015)}]{DurhamNotes}%
  \BibitemOpen
  \bibfield  {author} {\bibinfo {author} {\bibfnamefont {M.}~\bibnamefont
  {Jones}},\ }\href@noop {} {\enquote {\bibinfo {title} {Atom-light
  interactions},}\ }\bibinfo {howpublished} {Archived lecture notes, Durham
  University} (\bibinfo {year} {2015})\BibitemShut {NoStop}%
\bibitem [{\citenamefont {Benson}(2009)}]{HumboltNotes}%
  \BibitemOpen
  \bibfield  {author} {\bibinfo {author} {\bibfnamefont {O.}~\bibnamefont
  {Benson}},\ }\href
  {https://www.physik.hu-berlin.de/de/nano/lehre/copy_of_quantenoptik09/Chapter7}
  {\enquote {\bibinfo {title} {Interaction of atoms with a classical light
  field},}\ }\bibinfo {howpublished} {Archived lecture notes, Humbolt
  University Berlin} (\bibinfo {year} {2009})\BibitemShut {NoStop}%
\bibitem [{\citenamefont {Gr{\o}n}(1977)}]{Gron1977}%
  \BibitemOpen
  \bibfield  {author} {\bibinfo {author} {\bibfnamefont {{\O}.}~\bibnamefont
  {Gr{\o}n}},\ }\href {\doibase 10.1007/BF01811093} {\bibfield  {journal}
  {\bibinfo  {journal} {Int. J. Theor. Phys.}\ }\textbf {\bibinfo {volume}
  {16}},\ \bibinfo {pages} {603} (\bibinfo {year} {1977})}\BibitemShut
  {NoStop}%
\bibitem [{\citenamefont {Davies}\ \emph {et~al.}(1996)\citenamefont {Davies},
  \citenamefont {Dray},\ and\ \citenamefont {Manogue}}]{Davies1996}%
  \BibitemOpen
  \bibfield  {author} {\bibinfo {author} {\bibfnamefont {P.~C.~W.}\
  \bibnamefont {Davies}}, \bibinfo {author} {\bibfnamefont {T.}~\bibnamefont
  {Dray}}, \ and\ \bibinfo {author} {\bibfnamefont {C.~A.}\ \bibnamefont
  {Manogue}},\ }\href {\doibase 10.1103/PhysRevD.53.4382} {\bibfield  {journal}
  {\bibinfo  {journal} {Phys. Rev. D}\ }\textbf {\bibinfo {volume} {53}},\
  \bibinfo {pages} {4382} (\bibinfo {year} {1996})}\BibitemShut {NoStop}%
\bibitem [{\citenamefont {Bell}\ and\ \citenamefont
  {Leinaas}(1987)}]{Bell1987}%
  \BibitemOpen
  \bibfield  {author} {\bibinfo {author} {\bibfnamefont {J.}~\bibnamefont
  {Bell}}\ and\ \bibinfo {author} {\bibfnamefont {J.}~\bibnamefont {Leinaas}},\
  }\href {\doibase 10.1016/0550-3213(87)90047-2} {\bibfield  {journal}
  {\bibinfo  {journal} {Nucl. Phys. B}\ }\textbf {\bibinfo {volume} {284}},\
  \bibinfo {pages} {488} (\bibinfo {year} {1987})}\BibitemShut {NoStop}%
\bibitem [{\citenamefont {Jin}\ \emph {et~al.}(2014)\citenamefont {Jin},
  \citenamefont {Hu},\ and\ \citenamefont {Yu}}]{Jin2014}%
  \BibitemOpen
  \bibfield  {author} {\bibinfo {author} {\bibfnamefont {Y.}~\bibnamefont
  {Jin}}, \bibinfo {author} {\bibfnamefont {J.}~\bibnamefont {Hu}}, \ and\
  \bibinfo {author} {\bibfnamefont {H.}~\bibnamefont {Yu}},\ }\href {\doibase
  10.1103/PhysRevA.89.064101} {\bibfield  {journal} {\bibinfo  {journal} {Phys.
  Rev. A}\ }\textbf {\bibinfo {volume} {89}},\ \bibinfo {pages} {064101}
  (\bibinfo {year} {2014})}\BibitemShut {NoStop}%
\bibitem [{\citenamefont {Unruh}(1976)}]{Unruh1976}%
  \BibitemOpen
  \bibfield  {author} {\bibinfo {author} {\bibfnamefont {W.~G.}\ \bibnamefont
  {Unruh}},\ }\href {\doibase 10.1103/PhysRevD.14.870} {\bibfield  {journal}
  {\bibinfo  {journal} {Phys. Rev. D}\ }\textbf {\bibinfo {volume} {14}},\
  \bibinfo {pages} {870} (\bibinfo {year} {1976})}\BibitemShut {NoStop}%
\bibitem [{\citenamefont {Hnizdo}(2012)}]{Hnizdo2012}%
  \BibitemOpen
  \bibfield  {author} {\bibinfo {author} {\bibfnamefont {V.}~\bibnamefont
  {Hnizdo}},\ }\href {\doibase 10.1119/1.4712308} {\bibfield  {journal}
  {\bibinfo  {journal} {Am. J. Phys.}\ }\textbf {\bibinfo {volume} {80}},\
  \bibinfo {pages} {645} (\bibinfo {year} {2012})}\BibitemShut {NoStop}%
\bibitem [{\citenamefont {Sonnleitner}\ and\ \citenamefont
  {Barnett}(2017)}]{Sonnleitner2017}%
  \BibitemOpen
  \bibfield  {author} {\bibinfo {author} {\bibfnamefont {M.}~\bibnamefont
  {Sonnleitner}}\ and\ \bibinfo {author} {\bibfnamefont {S.~M.}\ \bibnamefont
  {Barnett}},\ }\href {\doibase 10.1140/epjd/e2017-80273-8} {\bibfield
  {journal} {\bibinfo  {journal} {Eur. Phys. J. D}\ }\textbf {\bibinfo {volume}
  {71}},\ \bibinfo {pages} {336} (\bibinfo {year} {2017})}\BibitemShut
  {NoStop}%
\end{thebibliography}

%

\end{document}